\documentclass[11pt]{article}
\usepackage{moriond,epsfig}

\def\be{\begin{equation}}
\def\ee{\end{equation}}
\def\bea{\begin{eqnarray}}
\def\eea{\end{eqnarray}}

\begin{document}
\vspace*{4cm}
\title{Heavy flavor physics at STAR}

\author{ J.Bielcik for STAR collaboration}

\address{Department of Physics, 
	        Yale University, 
 		New Haven, 06511, USA}

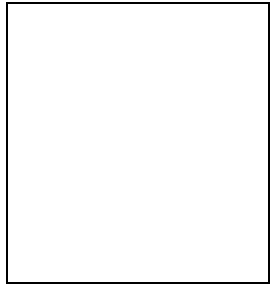
\maketitle\abstracts{
	In these proceedings, heavy flavor production as measured by STAR is disscussed. 
This can be done  directly by reconstruction of the  hadronic decays or indirectly by the  measurement of
electrons from semileptonic decays of heavy quark mesons. 
        The extracted charm production total cross-section per nucleon-nucleon collision in d+Au and Au+Au
collisions shows binary scaling, supporting the idea that charm quarks are produced in hard scattering in the initial 
phase of the collision.   
        The preliminary non-photonic electron spectra from $p+p$, $d+Au$ and $Au+Au$ collisions at
$\sqrt{s_{NN}}=200$ GeV at mid-rapidity are presented. The momentum range of reconstructed non-photonic electrons is  $1.5 < p_{T}< 10$ GeV/c.  The dominant contribution to the  non-photonic electron spectrum is the semi-leptonic decay of D and B mesons.  The electron nuclear modification factor ($R_{AA}$)  shows a large suppression in central Au+Au collisions,  indicating an unexpectedly  large energy loss for  heavy quarks in the hot and dense matter created at RHIC.  Theoretical models tend to overpredict the data if the contributions from both charm and beauty decays are taken into account.   }

\section{Introduction}

The measurement of inclusive hadron yields in central Au+Au collisions at RHIC 
led to the discovery of the  suppression of  hadron production at large transverse momenta ($p_{T}$)
compared to p+p collisions\cite{Star1,Adams:2003kv,Adams:2003im}.
 This is generally attributed to the energy loss of light partons 
in the dense nuclear matter created at RHIC. The energy loss itself depends on the properties of the medium, such as 
size and gluon density, as well as on the properties of the probe, such as color charge and mass. 
    Heavy quarks are believed to be mostly created through gluon fusion in the initial phase of the collision \cite{Lin:1995pk} and thus are excellent probes of the hot and dense matter.
 Due to the large mass of heavy quarks, 
the suppression of small angle gluon radiation should reduce their energy loss  and consequently, 
any suppression of  heavy-quark mesons at high $p_T$ is expected to be smaller than
that observed for hadrons consisting of light quarks\cite{Dokshitzer:2001zm}.
  
   Heavy quark mesons can be studied by the direct reconstruction of their hadronic decays, such as $D^0\rightarrow K^-\pi^+$. This direct 
reconstruction becomes a challenge  in Au+Au collisions, because of the very large combinatorial background. 
Therefore, the current measurements by STAR  are limited to $p_{T}<$ 3 GeV/$c$.
However, a combined fit, together with  low $p_{T}$ 
electron spectra\cite{STARcharmpaper,Zhang:2005} allows the $c-\bar{c}$ cross section to be measured.
The extracted values of the total charm cross-section are 1.33 $\pm$ 0.06(stat.) $\pm$ 0.18(sys.) mb
in 0-12$\%$ central Au+Au,  1.26$\pm$ 0.09 $\pm$ 0.23 mb in minimum bias Au+Au collisions and
1.4 $\pm$ 0.2 $\pm$ 0.2 mb in minimum bias d+Au collisions at $\sqrt{s_{NN}}=200$ GeV.
Within the systematic and statistical errors, the total charm production follows binary scaling from d+Au to Au+Au collisions, supporting the idea that heavy quarks are dominantly produced in the early stages of the collision in hard scatterings.

    An alternative way to infer information about  heavy quark production 
is the study of electrons from semi-leptonic decays of D and B mesons which is the focus of this paper. This method allows STAR to study charm and beauty production up to substantially larger $p_T$.  
There are several sources of  electrons that contribute to the inclusive spectra. We divide  them into two groups:
non-photonic electrons (signal) and photonic electrons (background). The non-photonic electrons are mainly 
from semi-leptonic decays of heavy mesons with a small contribution from  the Drell-Yan process. The  background photonic electrons are  from $\gamma$ conversions, and $\pi^0$, $\eta$ Dalitz and light vector-meson decays. 

\section{Electron identification  in STAR}

     The results presented in this paper were obtained from an analysis of 
data recorded with the STAR detector\cite{Ackermann:2002ad} in 2003 (p+p, d+Au) 
and  2004 (Au+Au). 
     Electron identification is based on a combination of energy loss
in the Time Projection Chamber, energy deposition in the Barrel Electro-Magnetic Calorimeter, 
and  shower profile in the Shower Maximum Detector. Further details of the analysis can be found elsewhere\cite{Bielcik:2005wu}. After all analysis cuts, a clean sample of electrons was obtained  with a $p_T$ dependent residual hadron 
contamination that varies from  10 to 15\%. 
    The data sample for the Au+Au dataset was divided into 3 centrality bins (0-5\%, 10-40\%, and 40-80\%). The electron reconstruction efficiency and acceptance  were determined by embedding simulated
electrons into real events. 
For the most central events, the electron reconstruction efficiency 
increases with $p_T$ up to  5~GeV/$c$ and then remains constant at 40\% .
    The photonic electrons have been statistically identified and
subtracted from the inclusive electron spectrum. The efficiency of the photonic background rejection was determined by embedding $\pi^0$, with a realistic $p_{T}$ distribution,    
into real events and is about 60\% for the most central Au+Au events, decreases slightly with $p_T$.

\section{Non-photonic electron spectra and $R_{AA}$}

\begin{figure}[h]
\centerline{\epsfxsize=10cm\epsfbox{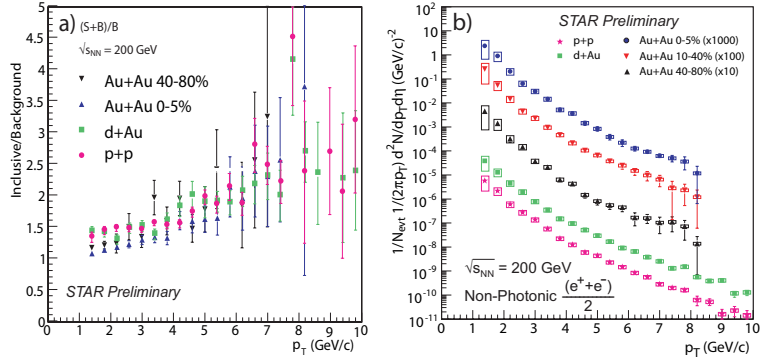}}   
\caption{ (a)Inclusive to photonic electrons ratio as a function of $p_T$. (b)Background subtracted non-photonic electron spectra for p+p, d+Au and Au+Au collisions with centralities 0-5\%, 10-40\% and 40-80\% . \label{fig:ele}}
\end{figure}

 In Figure~\ref{fig:ele}a, the ratio of inclusive electrons to photonic background electrons is shown as a function 
of  the $p_T$ of the electrons. For $p_T>$~2.0~GeV/$c$, there is a clear enhancement of electrons with respect to the background. This enhancement becomes more evident at higher momentum. Figure~\ref{fig:ele}b shows the preliminary background subtracted non-photonic electron spectra for p+p, d+Au and Au+Au collisions. The error  bars are statistical  and the boxes depict the preliminary systematic uncertainties.
 NLO pQCD \cite{Ramona:2005}, as well as Pythia LO pQCD calculations, predict that in a range between 3 and 6 GeV/$c$, the amount of electrons coming from B meson decays becomes dominant. STAR is capable of measuring non-photonic electrons in a momentum range above this transition. Although pQCD can predict the shape of the  non-photonic electron spectrum in p+p collisions rather well, it fails to describe the absolute electron yield by a factor $\approx$5. The shape of the spectrum can not be  described if only the contribution from charm decays would be taken into account.

\begin{figure}[h!]
  
\centerline{\epsfxsize=10cm\epsfbox{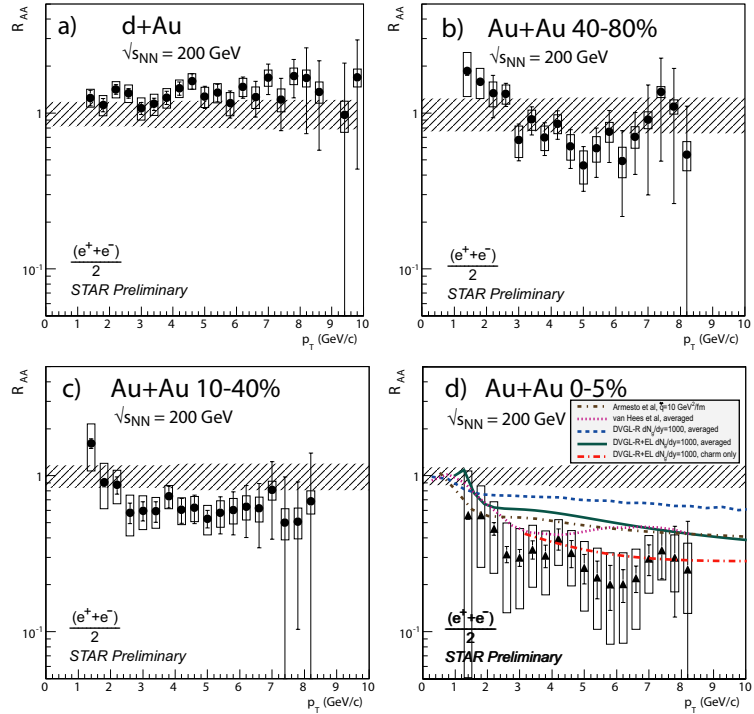}}   
\caption{Nuclear modification factors $R_{AA}$ for (a) d+Au, (b) Au+Au 40-80\%, (c) Au+Au 10-40\%, and (d) Au+Au 0-5\% comparison with models. \label{fig:raa}}
\end{figure}

Figure~\ref{fig:raa} shows the nuclear modification factors $R_{AA}$ and $R_{dAu}$ for non-photonic electrons as a 
function of $p_T$. The $R_{dAu}$ ratio (Figure~\ref{fig:raa}a) seems to be systematically above  unity for the entire $p_T$ range, 
consistent with a small Cronin enhancement. An increased suppression from peripheral to central Au+Au events 
(Figure~\ref{fig:raa}b-d) with 
respect to  binary scaling is observed. Assuming  that there is no other source of non-photonic  electrons, this suppression 
indicates a strong interaction and large energy loss of heavy quarks in the medium created at RHIC. Moreover, the amount of suppression in 
most central Au+Au collisions is similar to that of light hadrons. 

Figure~\ref{fig:raa}d also shows the  calculations of the  $R_{AA}$ from three theoretical models.
In all cases, the contribution from charm and bottom quarks were taken into account.  In the first 
model\cite{Armesto:2005mz} (dash-dotted curve), the  medium is characterized by the time averaged   BDMPS transport
 coefficient, $\hat q=10~GeV^2/$fm, for central Au+Au collisions. This value of $\hat q$ is consistent with the  
measurement of $R_{AA}$ of the light hadrons.  In the second 
model\cite{Djordjevic:2005} (dashed curve), the DGLV theory of radiative energy loss has been applied and  
the created nuclear matter is characterized by a gluon density of  $\frac{dN_g}{dy}$=1000, the value derived from 
light-quark meson suppression.  In addition, the
 contribution from elastic energy loss has been taken into account\cite{Wicks:2005gt} (solid line). For the sake of 
comparison, a  curve (dash-dot) representing the  contribution  only from charm sources is shown.
In the third model\cite{vanHees:2005wb} (dotted curve), the authors focus on elastic scattering of heavy quarks 
in the medium mediated by resonance excitations (D and B) off light quarks as well as by t-channel gluon exchange.
 
Although the data have large systematic and statistical errors, the data points tend to lie below the model calculations at high $p_T$. It is important to note  that the model calculations   also have large uncertainties, such as the amount of relative contribution from beauty/charm  decays, that influence the final $R_{AA}$. 
The FONLL calculation of the electron spectra from NLO pQCD \cite{Ramona:2005} depends also on   many parameters, such as
mass of the heavy quarks, $x_F$ (fragmentation scale) and $x_R$ (renormalization scale). Varying these parameters, one sees, that the $p_T$ range of the spectrum where beauty contribution starts to dominate over charm can be as low as  $\approx$3 GeV/c or as high as  $\approx$10 GeV/c.  Another issue is that the measured charm cross-section and also non-photonic electron spectra in p+p collions are about a factor $\approx$5-6 higher than FONLL predictions, although one would expect that due to the large mass of the  c and  b  quarks, pQCD should predict the spectra rather well. Also, a futher understanding of 
 p+p collisions is necessary before we can make a final statement about heavy quark energy loss.   

Recently the $R_{AA}$ of non-photonic electrons has been also calculated within the  Langevin model of heavy quark propagation 
in the created medium\cite{Teaney2005}. The propagation of the heavy quark in the medium is characterized by a diffusion coefficient, $D$. 
It was shown that electron $R_{AA}$ can be explained within this model if the value of  $D$ is between $12/(2\pi T)$ and $3/(2\pi T)$. Contributions from charm as well beauty semileptonic decays were taken into account.

\section{Conclusions}

      The non-photonic electron spectra measured by  STAR for p+p, d+Au, Au+Au collisions up to $p_T\approx$~8~GeV/$c$ were presented. An increasing suppression of non-photonic electrons with the collision centrality in Au+Au collisions is observed. This may be related to a stronger than predicted interaction between heavy quarks and the medium created at RHIC. The analysis of the full statistics from the 2004 Au+Au run will permit a more detailed study of the medium modifications for heavy quarks and  allow for a better understanding of quark energy loss mechanisms. 


\end{document}